Technical Report

# A Framework for Managing Evolving Information Resources on the Data Web

**Marios Meimaris**

m.meimaris@imis.athena-innovation.gr
Institute for the Management of Information Systems
Research Center Athena,
Greece

**George Papastefanatos**

gpapas@imis.athena-innovation.gr
Institute for the Management of Information Systems
Research Center Athena,
Greece

**Christos Pateritsas**

pater@imis.athena-innovation.gr
Institute for the Management of Information Systems
Research Center Athena,
Greece

**Theodora Galani**

theodora@imis.athena-innovation.gr
Institute for the Management of Information Systems
Research Center Athena,
Greece

**Yannis Stavrakas**

yannis@imis.athena-innovation.gr
Institute for the Management of Information Systems
Research Center Athena,
Greece

# A Framework for Managing Evolving Information Resources on the Data Web

Marios Meimaris, George Papastefanatos, Christos Pateritsas,
Theodora Galani, Yannis Stavrakas

Institute for the Management of Information Systems, Research Center "Athena", Greece
{m.meimaris, gpapas, pater, theodora, yannis}@imis.athena-innovation.gr

**Abstract.** The web of data has brought forth the need to preserve and sustain evolving information within linked datasets; however, a basic requirement of data preservation is the maintenance of the datasets' structural characteristics as well. As open data are often found using different and/or heterogeneous data models and schemata from one source to another, there is a need to reconcile these mismatches and provide common denominations of interpretation on a multitude of levels, in order to be able to preserve and manage the evolution of the generated resources. In this paper, we present a linked data approach for the preservation and archiving of open heterogeneous datasets that evolve through time, at both the structural and the semantic layer. We first propose a set of requirements for modelling evolving linked datasets. We then proceed on conceptualizing a modelling framework for evolving entities and place these in a 2x2 model space that consists of the semantic and the temporal dimensions.

## 1 Introduction

The emergence of the Data Web has brought forth a need to treat web information as more than a collection of linked documents, but rather as a dynamic accumulation of assertions and facts created within vibrant communities and collaborative environments, that can be processed, mined, combined and manipulated in a multitude of ways in order to extract and derive new knowledge. The use of implicit or explicit semantics and the linkage between autonomous and heterogeneous datasets attributes a larger value than the collection of standalone constituents yields.

    The Linked Data (LD) paradigm has shown the way towards publishing and interlinking data on the web, and appropriate standards and recommendations have ensured that this data is efficiently created, published and stored, as well as retrievable using standard query languages and frameworks (e.g. SPARQL, RDF). At the core lies that web resources are uniquely identified with the use of URIs and are interlinked in meaningful ways using typed links i.e., terms drawn from ontologies and vocabularies. This abstracts and unbinds resource representation from the functional interdependencies that come along with the technicalities of storing and publishing

the corresponding data. In this regard, publication and consumption of information are actions that become federated and non-uniquely controlled within their environment. The evolution of data objects can quickly become intractable if it is not captured and managed properly.

Considering the above, the benefits of evolution management can mainly be placed into two categories. The first one is concerned with quality control and maintenance. For instance, mishaps such as broken links or URI changes create inconsistencies and failures that are hardly feasible to overcome with no knowledge concerning the resources affected, their state of interconnectedness and the types of changes affecting them. The second one is concerned with data exploitation, as the evolution of data itself can provide valuable insights and shed light on the dynamics of the data, their domains and the operational aspects of the communities they are found in. In this sense, proper management of dataset dynamics in evolving open datasets addresses the following challenges [12]:

- Dataset synchronization: large chunks of data need to be replicated and maintained at external sources, with the need to be in constant sync with the original data source.
- Link maintenance: in LD, linksets are made of statements that link local resources to external resources. Sometimes external resources can disappear or change semantics, without necessarily taking appropriate steps to notify their dependents. This creates the need to identify such cases and take appropriate action.
- Vocabulary evolution and versioning: ontologies, vocabularies, data schemata etc. can change definitions and structure. This would have implications not only for their respective definitions, but for external instantiations of data that use these vocabularies as well. Therefore, such changes not only need to be communicated, but also have to be tackled accordingly in order to prevent loss of information and structure.
- Entity evolution. Information resources (entities) evolve across time and sometimes independently of their source dataset. Evolution management should address the evolution and changes of information entities at many levels of granularity in order to preserve the representation of entities across time.

Data-aware practices act as value drivers across domains [1][2], making persistence, long-term accessibility and usability of data essential value adding attributes. In this paper, we advocate the need for a well-designed framework for evolving datasets that addresses the problem in a multitude of dimensions. This framework should inevitably combine functionality such as versioning mechanisms, provenance tracking, change detection and quality control while at the same time provides efficient ways for querying the data both statically and across time. As a basis for the above, a thorough conceptual model that supports the representation of constructs relevant to the aforementioned dimensions and treats simple as well as complex changes as first-class citizens is required, and will be the main focus of this paper.

This paper is organized as follows. In section 2, we provide relevant works to preservation of data on the web and in section 3 we discuss on the requirements for handling evolving LOD datasets. Section 4 presents the evolution space model and its components, and section 5 describes the application of the model to real-world datasets. Finally, section 6 provides a conclusion and discusses future directions.

## 2   Related Work

Managing the evolution of linked open data is a problem of many dimensions. Multi-versioning, archiving, change detection and representation, model independence and provenance tracking are some of these. While work has been done in most of these fields individually, few approaches have regarded the issue as a singular problem of many interdependencies, less so in the case of the Data Web, where datasets evolve independently, often in non-centralized ways, while citing and using one another.

Relevant work for linked data versioning has been done in [3], where the writers propose an approach that builds on the Memento [4] framework, an effort to extend HTTP in order to include a temporal dimension that enables traversing through past versions of web documents, as well as querying them explicitly. In [3], a Memento-based approach is proposed for LD resources. At the representation level, the basic idea is to use a non-changing URI for current state identification and mint new URIs for each past version. The work focuses on individual LD resource versioning. Dereferencing a past version of the resource is done through temporal content negotiation, which is enabled by extending HTTP. Resources called 'Mementos' are used as access points for individual past versions, while the notion of a resource 'TimeGate' is used as a common access point for retrieving combined data of multiple past versions.

In [5], the authors tackle the problem of web versioning by providing extended functionality to the web server. They focus on web documents and components (HTML, images etc.). They associate versions with 'transaction times' and they perform the archiving process only when a web document is requested from the server. This creates a distinction between known and assumed past versions, making the whole process lossy and not consistent with realistic expectations for LD archiving. However, the system is not burdened with constant change tracking and makes a reasonable assumption that versions that have never been accessed are perhaps of no significant importance as far as archiving is concerned.

An interesting approach for managing and querying multi-version structured data is presented in [6]. The authors tackle the problem of multi-version management for XML documents in a similar way document management systems operate. They use deltas to capture the differences between two sequential document versions and use deltas as edit scripts to yield sequential versions. The introduced space redundancy is compensated by the intuitiveness of storing complete deltas (rather than compressed deltas) and its effects on query efficiency. They go on to define change detection as the computation of non-empty deltas and they argue that past version retrieval can be achieved by storing all complete deltas as well as a number of complete intermediate versions, finding the bounding versions of the desired ones and applying their corresponding deltas. Finally, they use a query language based on XQuery in order to enable longitudinal querying and they provide tag indices for each edit operation for faster delta application.

In [7], the authors propose a method for archiving scientific data from XML documents. The approach targets individual elements in the DOM tree of an XML document, rather than the whole versions themselves. They use time stamping in order to differentiate between the states of a particular element in different time intervals and they store each element only once in the archive. The timestamps are pushed down to the children of an element in order to reflect the changes at the corresponding level of the tree, an approach also followed in [8].

Interesting work has been done in [9] where the authors study the change frequency of LOD sources and the implications on dataset dynamics. They differentiate between the document-centric and the entity-centric perspectives of change dynamics, the latter further divided into the entity-per-document and global entity notions. We partially adopt this distinction in our work, as will be described further on.

SemVersion [10] computes the semantic differences as well as the structural differences between versions of the same graph but is limited to RDFS expressiveness. DSNotify [11] is an approach to deal with dataset dynamics in distributed LD. The authors identify several levels for the requirements of change dynamics: (1) vocabularies for describing dynamics, (2) vocabularies for representing changes, (3) protocols for change propagation and (4) algorithms and applications for change detection. It implements a change detection framework which incorporates these points in a unified functionality scheme, having as main motivation the problem of link maintenance. When dealing with changes, they target the what, how, when and why dimensions of the changes, closely related to the problem of provenance. They differentiate between triple and resource level for the what dimension and they argue that the level selection depends on the particular use case. How is expressed by the differential operators associated with a change (such as add/remove or compound changes) while when is expressed by timestamps and version numbers. Finally, the why dimension is usually expressed in manual annotations.

## 3 Requirements for evolving information resources

Most of the challenges related to the management of LD dynamics stem from the decentralized nature of publishing, curating and evolving interdependent datasets across multiple disparate sites. In contrast with traditional settings where data management and evolution is performed within a well-defined controlled environment where operations and dependencies on data can be easily monitored and handled, the Data Web poses new requirements for dataset evolution dynamics as follows:

**Persistent identification and reference (URIs, citations etc.)**: URIs lie at the core of the Linked Data paradigm as a means of entity identification between and across domains. Given that data published on the Data web are derived from different models and formats (e.g. relational data, csv files, etc.), there should be a uniform way of providing identifiers within the archiving framework reconciling the model-specific identification mechanisms across sources. Conceptually, this means that we have to provide an Identifier definition mechanism that will reify the information about the identifier itself as well as the identifier's nature (functional as well as non-functional metadata), e.g., primary keys must be converted to persistent URIs that represent citeable resources. In practice, we model this abstraction with the use of RDF URIs.

Another aspect involves the identification of entities across time. As new versions of such entities supersede older ones, it is important to differentiate between versions in order to be able to assert various kinds of metadata (temporal, provenance and so on) that describe these actions. This enables management and tracking of changes that take place on a certain entity in reference to other versions of itself. It is crucial, however, to maintain temporal persistence and identification on a higher level in order to provide a common reference for all instantiations (past and present) of the particular entity. This calls for differentiation between time-aware and time-agnostic representations of entities as well as appropriate time- and version-specific identification mechanisms.

**Simple and Complex Changes.** We consider changes as derivatives of comparing different versions of things, in varying degrees of representation. This implies that changes can be identified and asserted in a multitude of levels, depending on how semantic-rich and complex the information they capture is. For example, as is shown in [13], there is a hierarchical differentiation of changes that considers low-level changes such as additions and deletions as the essential building blocks for higher-level changes that in turn are more schema-specific and dependent on semantics and/or curator-based human interpretation. Building upon that, we consider that the ability of combining lower-level changes in order to describe higher-level ones is an essential requirement for providing rich semantics to any evolution management process. The higher the level of changes, the more context-dependent the issue becomes, as it is tied with factors such as the domain at hand, the design decisions, the underlying data, volume, dataset dynamicity and so on. These parameters come in combinations that create the need to study changes at different granularity levels. In order for changes to be detected and materialized on many levels, a formal and hierarchical representation model is required. Suitable change representation frameworks are expressive and extensible. This allows for defining axiomatic, mathematically complete low-level changes and at the same time provides a flexible mechanism for defining complex changes based on these.

Furthermore, it must be possible to define complex changes on higher semantic structures in order to give curators the ability to address and manage the change space in intuitive and dynamic ways [8]. Changes become domain-specific and case-relevant when they are properly manifested and instantiated. Our approach lies in the fact that it is not sufficient to identify what's changed, but how it is changed as well, and how it interacts with other changes and evolving attributes of identified subsets of the dataset at hand. However, the way an entity evolves is often found to be dependent on a subset of the data that is not always apparent, or directly attributed to the evolving entity explicitly. Furthermore, changes in themselves as conceptual constructs are not limited to the scope of just one identifiable entity. Rather, an instance of a change can affect a particular set of entities as a whole and still stand as a singularity of informational relevance to the dataset. This addresses the need for a formal way to define change contexts in datasets.

**Temporal and provenance annotations.** Provenance refers to all sources and catalysts of information (entities, processes, agents etc.) involved in producing or delivering an artefact. The term usually refers to the source or origin of an object, as well as the history, pedigree, lineage and passage through the object's various owners through time. The artefact can be anything from digital objects, to data stored in a database, to an object of the physical world. For this reason, provenance models and frameworks proposed in the literature are usually abstract enough so that they can be applied to various contexts and fields without much need for specialization.

In the scope of evolving datasets, provenance management enables or complements trust issues, interoperability between different sources, version control and licensing, among others. Provenance metadata can capture dataset lineage and this way provide added value to the data themselves. Provenance can be modelled in different granularities, from describing datasets as wholes to individual facts. It can generally be broken down to various categories in different dimensions, such as prospective vs. retrospective, work flow vs. data flow, fine-grained vs. coarse-grained, application-level vs. OS level and so on. In any case, artefact causality and informative annotations are generally considered as two of the main drivers in provenance modelling. The former refers to all sources, processes and origins relevant to the causes of an artefact's existence, while the latter refers to manually expressing the reasons for artefact creation/maintenance/modification and so on.

Within the context of evolving datasets, provenance is an issue of primary importance, as data from external sources are interlinked and reused from various other datasets and applications, thus creating the need to validate the origins of external data, as well as the processes and actors involved in the data's existence. We consider temporal metadata to be a part of provenance, and under this prism it can be argued that in the case of datasets, temporal provenance is a direct enabler of dataset versioning and consequently evolution. We adopt the partitioning of time into two categories, namely transaction-time and valid-time. It is within our belief that a data model for capturing evolution should exhibit these features and provide appropriate hooks in varying granularities, in order to track temporality at any needed level of detail.

**Common abstraction data model.** Considering the fact that linked and open data do not always follow prescribed directives and recommendations (e.g. RDF and SPARQL) and often use heterogeneous data models, including both standard formats and ad hoc or proprietary formats such as Excel spreadsheets or scientific data formats, problems arise when archiving evolving datasets of heterogeneous sources that are interlinked. Thus diachronic preservation should exhibit format-independence, data traceability and reproducibility, facilitation of evolution management services and an overall common denominator for data that originate from different models (e.g. ontological, relational, multidimensional etc.). In order for a framework to support the above, a uniform model is required to accommodate source, published, and archived datasets, contending with various forms of structured data alike. This will enable for interoperability between all sets of metadata (temporal, provenance, versioning etc.) that are associated with evolving datasets, and will provide a solid basis for query answering.

**Support for low-level (structural) and high-level (semantic & change-oriented) preservation.** Most importantly, the model must be able to capture the evolution of both the structural aspects of datasets (syntactic integrity) and the evolution of meaningfully defined information entities across time and versions. The former point ensures that the incoming datasets are reproducible at any point in the future using the same mechanisms independently of the source formats. The latter point enables for meaningful, high-level evolution queries to be answered, in ways that are not necessarily inherent to the dataset before it is archived. Therefore the model must retain the information as modelled on the datasets originally, and at the same time must take advantage of information-rich content and implicit or explicit semantics in incoming datasets and curator processes. The model provides a framework that can be used to define semantically richer entities on top of the semantics already depicted by the source datasets.

**Multi-versioning and longitudinal querying.** A basic requirement for long-term preservation and evolution management is to support multiple versions of entities as well as querying the history of them. Furthermore, proper representation and exposition techniques that allow for citing time-agnostic entities and particular versions of entities should be supported as well. This provides a way to explicitly talk about or link to past versions and helps avoid the problem of broken links when interlinking data from external datasets. Such a framework should be able to answer a number of different types of queries, ranging from the dataset metadata, changes and other evolutionary aspects, as well as the data themselves. Furthermore, in data with temporal relevance, we consider time to define two types of temporal queries, namely snapshot queries (fixed-time or version) and longitudinal queries (queries across different versions or temporal periods). Some typical types of queries the framework should address are:

- *Dataset listing*: Retrieve a list of datasets stored in the archive. The list can either be exhaustive or filtered based on sets of criteria.
- *Dataset versions listing queries*: Retrieve a list of the available versions of a given dataset. The list can either be exhaustive or filtered based on temporal, provenance or other metadata criteria.
- *Complete dataset queries*: Retrieve version(s) of a dataset, filtered accordingly (e.g. specific version identifier, temporal criteria, etc.).
- *Partial dataset queries* (data views): Retrieve part(s) of a dataset. The parts can be predefined based on the archive data model (described in the next section), or dynamically isolated based on appropriate declarative criteria.
- *Longitudinal queries*: As above but with the complete timeline of the data (Partial dataset queries). Temporal criteria can be applied to limit the timeline (specific versions or time periods), or successive versions.
- *Queries on Changes*: Retrieve changes between two concurrent versions of an entity (dataset, resource etc.). Limit results for specific type of changes, or for a specific part of the data (see partial dataset queries).
- *Mixed Queries on Changes and Data*: Retrieve datasets / parts of datasets that are affected by specific types of changes.

## 4      Modelling Evolution: the 2x2 Model Space

Our approach is depicted in figure 2, which shows the division of the evolution model space in a 2x2 grid. The vertical dimension divides the space into time-agnostic and time-aware; a distinction between entities that evolve over time on the bottom half of the grid, and entities that are unaware of the temporal aspect in the top half. The horizontal dimension divides the space into the raw data space and the curated information space. The data space holds data and structure information of the archived datasets without any further semantics than what is already attributed to them at the time of deployment, whereas the curated information space contains subsets of the data space, explicitly defined by the curator of the archive, and with the purpose of capturing the evolution of specific information resources. At the core of the model lies the notion of the evolving entity, which captures both structural and semantic constructs of a dataset and acts as a common placeholder for provenance, temporal, and other types of metadata (Fig 1). Evolving entities are identifiable and citable objects. In this sense, the following types of objects are considered as entities:

- Diachronic Datasets (time-agnostic entity that represents a dataset)
- Dataset instantiations (time-specific instantiation of a diachronic dataset)
- Schema Objects (e.g. classes and properties in ontologies, tables and columns in relational databases and so on)
- Data Objects (entities that represent data entries for heterogeneous models)
- Diachronic Resources (e.g. ontology instances, linked data resources etc.)

These entities are modeled in such a way that their basic information is captured at the following three levels, as can be seen in Figure 1:
- Identification
- Attributes (type-specific information related to an entity)
- Provenance

This offers the advantage of facilitating the aforementioned requirements on many different levels, as well as using the same mechanisms in change detection and provenance tracking independently of what is being monitored, thus providing fine granularity at different syntactic and semantic levels. Every type of entity can hold its respective provenance metadata; every entity can be compared to entities of the same type for changes, not only across datasets, but across different versions of the same dataset as well.

In any case, there should be a uniform way of providing identifiers, given that entities are identified in different ways across formats (e.g. resource URIs vs. primary keys, property URIs vs. column names). For this reason, a level of abstraction is needed in order to bring the identifiers to the same level. One way to achieve this is to provide an Identifier concept that will reify the information about the identifier itself as well as the identifier's nature (functional as well as non-functional metadata). Furthermore, using this level of abstraction the identifiable entity can be associated not only with its identifier(s), but with metadata such as temporality, provenance and so on.

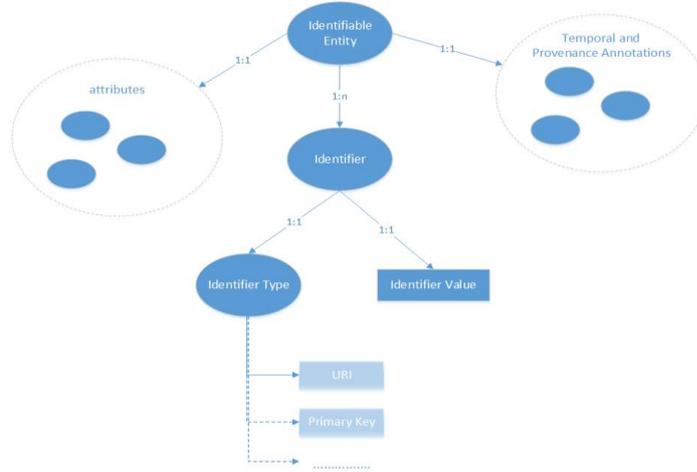

**Fig. 1.** Evolving, identifiable entities

The main types of evolving entities comprising the proposed model are described as following.

**Diachronic datasets and dataset instantiations.** Diachronic datasets are conceptual entities that represent a particular dataset from a time-agnostic point of view. Different temporal instantiations of the same diachronic dataset are linked to the latter in the data model. Furthermore, diachronic dataset metadata comprise information that is not subject to change, such as diachronic dataset identifiers. These identifiers serve as ways to refer to the datasets in a time and/or version unaware fashion (i.e. diachronic citations). Dataset instantiations hold information on how and when a particular dataset instantiation was relevant, active, trusted and so on. In essence they define (temporal) versions of diachronic datasets. These instantiations do not contain their records; rather they are linked to them.

**Record sets**. Record sets are, collections of records that exist within a particular dataset instantiation. Given a particular record set and the dataset's metadata information, the dataset instantiation can be recalled and reproduced in its original form. Furthermore, past instantiations of datasets can be interpreted using different schemata, not only the one(s) relevant to the time the instantiation was current.

**Schema Objects.** Schema objects represent the schema-related entities of the archived datasets given the dataset's source model. For instance, the classes along with their class restrictions of an ontology, the properties and their definitions (domains, ranges, meta properties depending on the expressivity) are modelled as schema objects. The goal is to provide a reusable modelling mechanism for identifying and referring to schema elements and their evolution across datasets. In this way, schema evolution can be captured, by annotating schema elements with evolution changes. Furthermore, schema versions can be applied on datasets independently of whether they co-existed in a dataset instantiation or not.

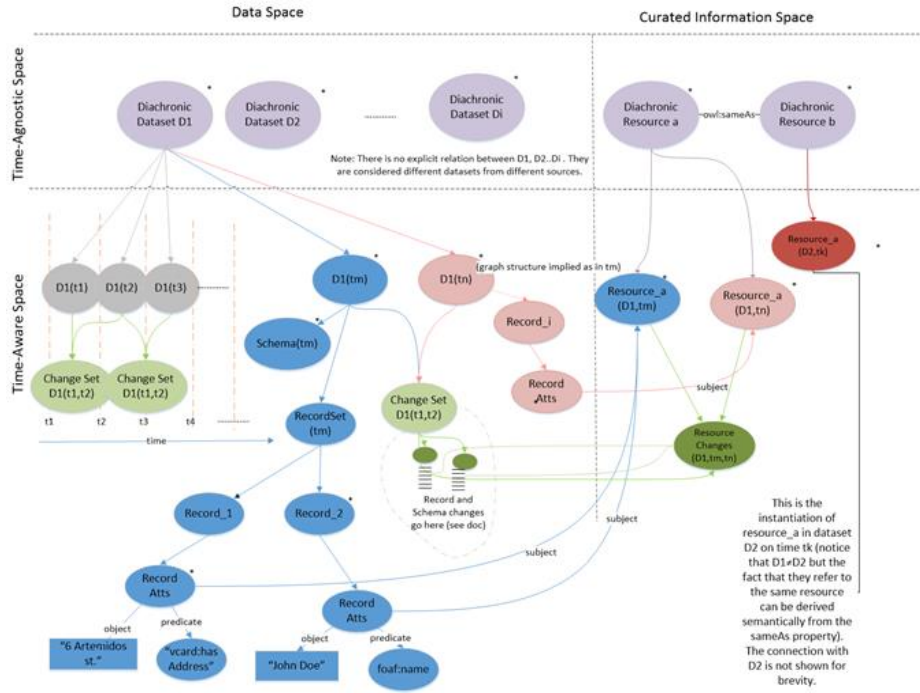

**Fig. 2.** The 2x2 Model Space

**Data Objects (Records and Record Attributes).** Data objects consist of records and record attributes. A record represents a most granular data entry about a particular evolving entity. Records are uniquely identified in order to make record-level annotation feasible and their type is characterized by the type of the source dataset populating the archive (e.g. tuples for relational datasets, observations for multidimensional, triples for RDF, etc.) in order to attribute provenance, temporality and changes on them. Given that a record can be comprised of one or more binary relationships, depending on the original format, it serves as a bag holder of one or more record attributes. The approach followed is that every data record (tuple, row etc.) is broken down to assertions (facts) that can be expressed with triples of the form <subject> <predicate> <object>. In this sense, a record reifies the predicate-object pairs for a fixed subject. These predicate-object pairs are called record attributes. For instance, a tuple from a relational table is considered to be a record describing the tuple's primary key, with each relational attribute being a record attribute.

**Diachronic Resources and resource instantiations.** Resources are collections of user-defined identifiable concepts within the archiving framework defined through a declarative way, i.e., a view, over a dataset's records. They essentially provide a curation mechanism to define contexts (i.e. parts of the dataset) of evolution and relate high level changes to them. Similarly to diachronic datasets, a diachronic resource provides a persistent view on the semantics of the curated resource, which is not subject to change, such as diachronic resource identifiers. The resource instantiation captures the resource evolution across time and its realization over a versioned dataset's records. The definition of a resource consists of two parts; the resource identification definition gives the way an instantiated resource is identified within the archive. The resource description definition provides the way a resource is evaluated over the records of a particular dataset instantiation. This process outputs the evolving context of a resource (fig 2).

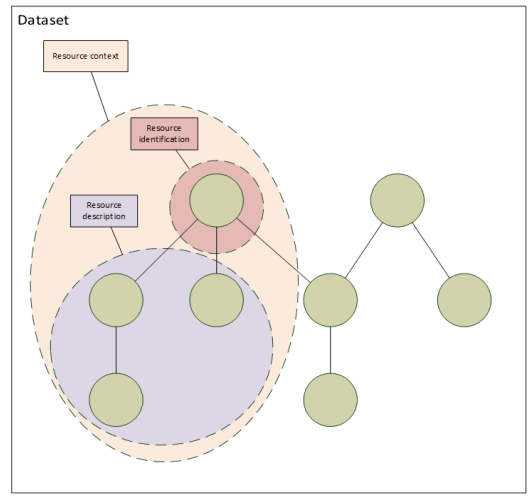

**Fig. 3.** A resource context, made up from resource identification and description

Resources can be versatile in nature across datasets and data formats. For example, given an ontology and its instantiation, each class instance can describe a resource identified by the respective URI. Given a table of employees in a relational database, a resource in this sense can be a particular employee identified by his primary key. Finally, in a multidimensional dataset, a resource can be a specific observation identified by the values of the constituent dimensions. More complex definitions of resources are allowed and, in fact, encouraged for capturing more high-level, curator specific semantics of evolution and dataset dynamics. For instance, a diachronic resource can be identified by more than one subject (e.g. defining a diachronic resource as the set of n particular employees) and limits can be imposed on the definition of the resource's description (e.g. defining a diachronic resource's description as a subset of its subjects' triples, and some second-level triples of the objects that appear in the first-level triples of the subjects).

**Changes and change sets.** Changes are only relevant to pairs of two diachronic entities of the same type (e.g. changes between two diachronic resources, two dataset instantiations, two schema objects etc.). In a time-oriented versioning scenario, changes come in Change Sets between two dataset instantiations of the same diachronic dataset. These are comprised of change entities that refer to changes between particular entity types (e.g. changes between record sets, changes between schemata and so on) in the two datasets under comparison.

## 5      Verifying the 2x2 Model Space over real world datasets

In this section, we verify the proposed model over three different real world datasets coming from heterogeneous application domains, namely (i) an open data application employing statistical and governmental multidimensional data; (ii) a biological application manipulating experimental results modelled as RDF linked data with custom ontologies and finally (iii) an enterprise application drawn from the automotive industry using relational data. At the dataset level, the process prescribes that datasets are given a diachronic identifier, and linked to each of their temporal versions. Change detection takes place between two dataset versions in order to compute the low-level (i.e. deltas) and high-level (i.e. simple and complex changes) between them. These are used to relate versions of evolving entities and be able to retrieve evolution information by treating changes as part of the model.

Our model abstracts data entries to records and record attributes. A record attribute represents a very fine-grained fact that essentially gives a value to a property. Record attributes are associated with records that in turn serve to aggregate all facts about a certain entity. In a relational model, rows are mapped to records and fields are mapped to record attributes. In the multidimensional model, we treat multidimensional observations as records, and their dimensions and measures as record attributes. In the ontological model, we consider a record as the placeholder of all data describing a particular entity. For instance, in RDF, the relationship between a record and a subject URI is 1:1, and all predicate-object pairs for this particular subject are linked to this record. In other words, each URI described in a dataset creates its own record, which in turn is linked to a number of record attributes equal to the number of triples the URI is the subject of. In following, we propose a set of rules to map such models to the 2x2 space model. For each time-agnostic representation of a dataset, a corresponding Diachronic Dataset entity has to be created. Each time-specific version of the dataset is therefore linked to the diachronic dataset.

**Mapping relational datasets.** The approach is closely aligned to the R2RML[1] W3C recommendation, where tables are used to generate classes and columns are used to generate properties. The mapping rules for relational data are as follows:

1. For each dataset version, relations are mapped to classes and columns are mapped to properties, which are both schema objects. The created properties are assigned with a domain that corresponds to the relation class and a range for the corresponding SQL type (e.g., double, string, etc.).
2. Primary (Simple or Composite) Keys are identified and given URIs. The PK values are iterated in order to create resources that will act as subjects in the mapped triples. The created resources are typed via the rdf:type to the class of the relation that was created in the previous step.

---
[1] http://www.w3.org/TR/r2rml/

3. Each tuple creates a record that is linked to the subject URI from (2). Each field value is mapped to a record attribute by using the property created in (1) as predicate and the actual value as object. These record attributes are linked to the record.

**Mapping multidimensional datasets.** Multidimensional models are modeled as profiles of the Data Cube Vocabulary[2]. The latter defines the concepts of observations, dimensions, measures and attributes. Observations bear close resemblance to records, as they essentially represent data entries that describe how a set of dimension values is expressed in a particular measurement. Dimensions, measures and attributes are properties that observations hold. Considering the above, the guidelines for mapping multi-dimensional data to our model are as follows:

1. Measures, dimensions and attributes are mapped to properties as in the case of column in relational data.
2. Each observation creates a record and is linked with record attributes that are connected with the corresponding dimension, measure and attribute properties.

**Mapping ontological datasets.** Ontological \ RDF data are mapped to the proposed model in a most straightforward manner:

1. RDF classes, the properties and their definitions (domains, ranges) are modelled as schema objects.
2. The RDF triples are grouped by their subjects. For each subject URI, its corresponding collection of triples (i.e. triples that feature this URI as a subject) are grouped in record. This record is connected with the record attributes created for each triple associated with this URI as a subject.

## 6   Conclusions and Future Work

In this paper, we have presented our efforts towards a framework for managing evolving information resources on the Data Web, on many granular levels. We have first presented the challenges and requirements for preservation and evolution management of heterogeneous web datasets. We went on to argue that a common abstraction model is needed in order to maintain the structural as well as the semantic aspects of the data and we have introduced the notion of diachronic resources as ways to define contexts for high-level changes. We have proposed a model for evolution, the components of which can reside into a 2X2 space where objects are separated by their temporal dependence and their curator-imposed evolution semantics and showed how to map real-world scenarios to the model.

The work deals with the conceptual characteristics of the issue and does not deal with technical issues of implementation, such as archiving structures and strategies. Our next work is to employ this model for implementing an archiving framework for evolution management of open datasets and address issues of model serialization, scalability and performance.

---

[2] http://www.w3.org/TR/2014/REC-vocab-data-cube-20140116/